\documentclass[prb,twocolumn,showpacs]{revtex4}
\usepackage{amssymb}
\usepackage{amsmath}
\usepackage{graphicx}
\usepackage{bm}
\usepackage{color}
\usepackage{ulem}
\usepackage[T2A]{fontenc}
\newcommand{\braket}[3]{\left\langle #1 \left| #2 \right| #3 \right\rangle}
\newcommand{\ket}[1]{\left| #1 \right\rangle}
\newcommand{\dz}[0]{\frac{d}{dz}}

\DeclareMathAlphabet{\mathantt}{OT1}{antt}{li}{it}

\begin{document}

\title{Spin-orbit splitting of valence subbands in semiconductor nanostructures}
\author{M. V. Durnev}
\author{M. M. Glazov}
\author{E. L. Ivchenko}
\affiliation{Ioffe Physical-Technical Institute of the RAS, 194021 St. Petersburg, Russia}

\begin{abstract}
We propose the 14-band $\bm k \cdot \bm p$ model to calculate spin-orbit splittings of the valence subbands in semiconductor quantum wells. The reduced symmetry of quantum well interfaces is incorporated by means of additional terms in the boundary conditions which mix the $\Gamma_{15}$ conduction and valence Bloch functions at the interfaces. It is demonstrated that the interface-induced effect makes the dominating contribution to the heavy-hole spin splitting. A simple analytical expression for the interface contribution is derived. In contrast to the 4$\times$4 effective Hamiltonian model, where the problem of treating the $V_z k_z^3$ term seems to be unsolvable, the 14-band model naturally avoids and overcomes this problem.  Our results are in agreement with the recent atomistic calculations [J.-W. Luo  et al., Phys. Rev. Lett. {\bf 104}, 066405 (2010)].
\end{abstract}

\maketitle

%%%%%%%%%%%%%%%%%%%%%%%%%%%%%%%%%
\section{Introduction}
%%%%%%%%%%%%%%%%%%%%%%%%%%%%%%%%%
As follows from the time inversion symmetry and the Kramers theorem, the electronic states in centrosymmetric systems are at least doubly-degenerate. On the other hand, in the three-, two- and one-dimensional systems lacking a center of space inversion, the degeneracy of free Bloch single-electron states is removed with exception of particular points and directions of the Brillouin zone. The removal of spin degeneracy occurs due to spatially antisymmetric part of the single-particle periodic Hamiltonian ${\cal H}({\bm r}, \hat{\bm p})$  
with allowance for the spin-orbit interaction. In terms of the effective Hamiltonian ${\cal H}({\bm k})$ the interaction appears as a spin-dependent contribution odd in the electron wave vector ${\bm k}$. This contribution is responsible for a number of fascinating and important effects being actively studied nowadays, see, e.g., Refs.~[\onlinecite{opt_or_book,Fabian04,Fabian07,dyakonov_book,SST2008,SST2009,Semicond,wu_review_2010}].

In nano- and heterostructures, the quantum confinement strongly modifies free-carrier dispersion. Particularly in quantum wells (QWs), the parabolic conduction band turns into series of two-dimensional subbands shifted, in parallel, along the energy axis. In the absence of inversion center, the spin-orbit interaction splits each subband with the splitting described by linear and, sometimes, cubic ${\bm k}$ terms in the 2$\times$2 effective Hamiltonian.\cite{winkler:book,Ivchenko} Because of the more complex valence band structure, the dispersion of holes is also much more complicated than that in the conduction band. Rashba and Sherman \cite{Rashba1988175} were the first to calculate the spin splitting of the topmost heavy- and light-hole subbands in QWs grown along $z\parallel [001]$ from zinc-blende lattice semiconductors by using the bulk effective Hamiltonian (four-by-four matrix) consisting of the conventional Luttinger Hamiltonian and spin-dependent terms of the order of ${\bm k}^3$. They imposed the simplest conditions for the four-component hole envelope wave function, $\psi=0$, on the boundaries of the QW and obtained $\bm k$-linear terms in the effective Hamiltonians of two-dimensional hole subbands. Direct extension of the procedure developed in Ref.~\onlinecite{Rashba1988175} for the realistic models of quantum confinement, particularly, including the effects of finite barrier, deems impossible owing to the presence of $k_z^3$ spin-dependent term in the bulk $4\times 4$ Hamiltonian. This term makes consistent matching of the valence-band wave functions really challenging, since it leads to a $k_z^2$ contribution to the velocity operator $\hat{v}_z$ and, therefore, to a singularity of the flux $\propto \hat{v}_z \psi$ at the interface.

Here we develop the $14$-band ${\bm k}\cdot{\bm p}$ model to calculate spin splittings of hole subbands in QWs, which allows us to avoid the $k_z^3$-term problem. Moreover, we propose additional terms in the boundary conditions for the 14-component envelope which naturally describe the interface heavy-light hole mixing arising due to anisotropy of chemical bonds at the interfaces.~\cite{aleiner92eng,ivchenko96,krebs96,Vervoort,Guettler,Toropov} The developed ${\bm k}\cdot{\bm p}$ approach presents an independent alternative to atomistic calculations of the spin-orbit splittings in QWs.~\cite{Foreman,PhysRevLett.104.066405}

The paper is organized as follows: Sec.~\ref{sec:sym} presents the symmetry analysis of the spin-dependent terms in the valence band Hamiltonian, Sec.~\ref{sec:14} outlines the 14-band $\bm k\cdot \bm p$ model and the spin-orbit splittings in the bulk semiconductor as well as boundary conditions for the QW structures; numerical results and analytical approximations are presented in Sec.~\ref{sec:res}, and Sec.~\ref{sec:concl} contains brief conclusions.

%%%%%%%%%%%%%%%%%%%%%%%%%%%%%%%%%
\section{Symmetry considerations}\label{sec:sym}
%%%%%%%%%%%%%%%%%%%%%%%%%%%%%%%%%
\subsection{Bulk zinc-blende-lattice crystals}
%%%%%%%%%%%%%%%%%%%%%%%%%%%%%%%%%
We begin the symmetry analysis by reminding that, in a bulk zinc-blende-lattice semiconductor, the expansion of spin-dependent part 
${\cal H}_c^{(3)}$ of the electron effective Hamiltonian in the conduction band $\Gamma_6$ starts from the nonzero cubic term 
\begin{equation} \label{eq:cond}
{\cal H}_c^{(3)} = \gamma_c \left( {\bm \sigma}{\bm \varkappa} \right),
\end{equation}
where $\gamma_c$ is the band parameter, $\bm \sigma$ and $\bm \varkappa$ are pseudovectors composed of the Pauli matrices in the coordinate system $x\parallel [100]$, $y\parallel [010]$, $z\parallel [001]$ and the cubic combinations $\varkappa_x = k_x (k_y^2 - k_z^2)$ etc. The expansion of the 4$\times$4 effective Hamiltonian in the $\Gamma_8$ valence band starts from the first order term ${\cal H}_v^{(1)} = (4${\small{$\mathantt{k}_0$}}/$\sqrt{3}) {\bm V} {\bm k}$, where $V_x = \{J_x, J_y^2 - J_z^2\}_s$, $\{A,B\}_s = (AB + BA)/2$, $J_{\alpha}$ are the angular momentum matrices in the basis of spherical harmonics $Y_{3/2,m}$. \cite{Pikus1988,ivchenkopikus} 
Although the T$_d$ point symmetry allows this term, the coefficient {\small{$\mathantt{k}_0$}} is nonzero only due to the ${\bm k}\cdot{\bm p}$ mixing between the valence states and the remote $\Gamma_3$ states, it is quite small and may be neglected for the GaAs-based systems.\cite{Pikus1988,ivchenkopikus} In the $\Gamma_8$ band, the cubic-${\bm k}$ term contains three linearly independent contributions, as follows, 
\begin{multline} \label{eq:G8SS}
{{\cal H}^{(3)}_{v}  = \gamma_v ({\bm J}{\bm \varkappa} )}\\
 +\frac{1}{2} \delta \gamma_v 
\left[ a_2 \sum_{\alpha} J_{\alpha}^3 \varkappa_{\alpha} + a_3 \sum_{\alpha} V_{\alpha} k_{\alpha} \left(k_{\alpha}^2 - \frac13 k^2 \right) \right]\:, 
\end{multline}
where, for further convenience, two of the three coefficients are presented as products of the parameter $\delta \gamma_v$ which has the dimension of $\gamma_c, \gamma_v$ and dimensionless factors $a_2/2$ and $a_3/2$.\cite{Pikus1988} It is noteworthy that the first term is non-relativistic in its origin, it is symmetry-allowed for the $\Gamma_{15}$ band. Two terms in the second line of Eq.~\eqref{eq:G8SS} are relativistic and one can see that the last summand contains the ``dangerous'' contribution proportional to $k_z^3$.
%%%%%%%%%%%%%%%%%%%%%%%%%%%%%%%%%%%%%%%%%%%%%%%%%
\subsection{Quantum well structures}
%%%%%%%%%%%%%%%%%%%%%%%%%%%%%%%%%%%%%%%%%%%%%%%%%
In the following symmetry analysis we consider only symmetrical QW structures grown along the crystallographic axes [001], 
[110] or [111] and having the point symmetries (i) D$_{2d}$, (ii) C$_{2v}$ and (iii) C$_{3v}$, respectively. The Cartesian coordinates are conveniently chosen along the axes (i) $x \parallel [100],\: y \parallel [010],\: z \parallel [001]$ or $x_1 \parallel [1\bar{1}0],\: y_1 \parallel [110],\: z_1 \parallel [001]$, (ii) $x_2 \parallel [1\bar{1}0],\: y_2 \parallel [001],\: z_2 \parallel [110]$, and (iii) $x_3 \parallel [11\bar{2}],\: y_3 \parallel [\bar{1}10],\: z_3 \parallel [111]$.
In a QW structure the $\Gamma_8$ valence band is split into the heavy- and light-hole-like states. In the following we choose the basic states at the $\Gamma$-point ($k_{x_j} = k_{y_j} = 0$) transforming under the symmetry operations as the Bloch functions
\begin{eqnarray} \label{psihh}
&&\psi_{hh}^{(1)} \equiv |\Gamma_8, - 3/2 \rangle = \downarrow (\mathcal X_j - {\rm i}\mathcal Y_j)/\sqrt{2},  \\ 
&&\psi_{hh}^{(2)} \equiv |\Gamma_8, 3/2 \rangle = - \uparrow (\mathcal X_j + {\rm i} \mathcal Y_j)/\sqrt{2} \nonumber
\end{eqnarray}
or 
\begin{eqnarray} \label{psilh}
&&\psi_{lh}^{(1)} \equiv |\Gamma_8, 1/2 \rangle = [2 \uparrow \mathcal Z_j - \downarrow (\mathcal X_j + {\rm i}\mathcal Y_j)]/\sqrt{6},  \\ 
&&\psi_{lh}^{(2)} \equiv |\Gamma_8, -1/2 \rangle = [2 \downarrow \mathcal Z_j + \uparrow (\mathcal X_j - {\rm i} \mathcal Y_j)]/\sqrt{6}\:. \nonumber
\end{eqnarray} 
Here $\uparrow, \downarrow$ are the spin-up and spin-down two-component columns, $\mathcal X_j, \mathcal Y_j$ and $\mathcal Z_j$ are the periodic orbital Bloch functions in the chosen coordinate system $x_j, y_j, z_j$ ($j= 1,2,3$), and for simplicity we omit the index $j$ in the spin columns.  

\subsubsection{Growth direction [001]} 
The states in the heavy-hole subbands $hh1, hh3\dots$ and light-hole subbands $lh2, lh4\dots$ transform according to the $\Gamma_6$ spinor representation of the point group D$_{2d}$ whereas the eigenstates $lh1, lh3\dots$ and $hh2, hh4\dots$ form the bases for the $\Gamma_7$ representation. In the method of invariants\cite{ivchenkopikus}, the 2$\times$2 matrix effective Hamiltonian in each hole subband is decomposed into a linear combination of four basis matrices. Since both direct products $\Gamma_6 \times \Gamma_6$ and $\Gamma_7 \times \Gamma_7$ reduce to the same direct sum of irreducible representations $\Gamma_1 + \Gamma_2 + \Gamma_5$, the basis matrices can be chosen common for the subbands of $\Gamma_6$ and $\Gamma_7$ symmetries. If the basic functions of the spinor representations are chosen in the form~\eqref{psihh}, \eqref{psilh}, then the set of basic matrices includes the identity matrix ($\Gamma_1$ representation), pseudospin matrix $\sigma_{z_1}$ ($\Gamma_2$ representation) and two pseudospin matrices $\sigma_{x_1}, \sigma_{y_1}$ transforming as the pair of wave vector components $k_{y_1}, k_{x_1}$ ($\Gamma_5$ representation). This allows one to write the linear-${\bm k}$ term in the effective Hamiltonian as
\begin{equation} \label{100}
\mathcal H^{[001]}_n = \beta^{(n)}_1 (\sigma_{x_1} k_{y_1} + \sigma_{y_1} k_{x_1}) =\beta^{(n)}_1 (\sigma_{x} k_{x} - \sigma_{y} k_{y}) \:,
\end{equation}
where $n = hh \nu$ or $lh \nu$, $\nu = 1,2\dots$, and $\beta^{(n)}_1$ are the subband parameter. In Eq.~\eqref{100}, the effective Hamiltonian is presented in the two coordinate systems $x,y,z$ and $x_1, y_1,z_1$ relevant for the $(001)$ structures. We stress that the same form of the effective Hamiltonian for the heavy-hole and light-hole subbands results from the special order of the Bloch functions in Eqs.~\eqref{psihh} and \eqref{psilh}.
%%%%%%%%%%%%%%%%%%%%%%%%%%%%%%%%%%%%%%
\subsubsection{Growth direction [110]}
%%%%%%%%%%%%%%%%%%%%%%%%%%%%%%%%%%%%%%
Both heavy- and light-hole states transform according to the same spinor representation $\Gamma_5$ of the group C$_{2v}$. Among components $k_{x_2}, k_{y_2}$, $\sigma_{\alpha_j}$ only $k_{x_2}$ and $\sigma_{z_2}$ transform according to equivalent representations. As a result, the linear-${\bm k}$ term has the form
\begin{equation} \label{110}
\mathcal H^{[110]}_n = \beta^{(n)}_2 \sigma_{z_2} k_{x_2}\:,
\end{equation}
with $\beta^{(n)}_2$ being the subband parameters.
%%%%%%%%%%%%%%%%%%%%%%%%%%%%%%%%%%%%%%
\subsubsection{Growth direction [111]}
%%%%%%%%%%%%%%%%%%%%%%%%%%%%%%%%%%%%%%
The pair of functions (\ref{psihh}) and the states $hh \nu$ transform according to the reducible representation $\Gamma_5 + \Gamma_6$ of the C$_{3v}$ point group. The direct product $(\Gamma_5 + \Gamma_6) \times (\Gamma_5 + \Gamma_6) = 2\Gamma_1 + 2\Gamma_2$ does not contain the $\Gamma_3$ representation which means that the $\bm k$-linear splitting of the heavy-hole states is symmetry-forbidden. The first non-vanishing spin-dependent contribution to the heavy-hole effective Hamiltonian is cubic in ${\bm k}$ and has the form~\cite{wu}
\begin{eqnarray} \label{hh111}
\mathcal H_{hh \nu}^{[111]} = \gamma_1^{(\nu)} \sigma_{x_3} k_{y_3} \left( k_{y_3}^2 - 3k_{x_3}^2 \right) \hspace{1 cm} \\
+ \gamma_2^{(\nu)} \sigma_{y_3} k_{x_3} \left( k_{x_3}^2 - 3k_{y_3}^2 \right) + \gamma_3^{(\nu)} \sigma_{z_3} k_{y_3} \left( k_{y_3}^2 - 3k_{x_3}^2 \right) \nonumber
\end{eqnarray}
with three independent parameters $\gamma_1^{(\nu)}$, $\gamma_2^{(\nu)}$ and $\gamma_3^{(\nu)}$.

By contrast, $\bm k$-linear terms are allowed in the dispersion of light-hole subbands. Indeed, the functions (\ref{psilh}) and the light-hole states $lh \nu$ transform according to the two-dimensional representation $\Gamma_4$. The product $\Gamma_4 \times \Gamma_4 = \Gamma_1 + \Gamma_2 + \Gamma_3$ contains a $\Gamma_3$ representation meaning that the $\bm k$-linear light-hole splitting is described by 
\begin{equation} \label{lh111}
\mathcal H_{lh \nu}^{[111]} = \beta_3^{(\nu)} \left( \sigma_{x_3} k_{y_3} - \sigma_{y_3} k_{x_3} \right)\:,
\end{equation}
with a single parameter $\beta_3^{(\nu)}$.

%%%%%%%%%%%%%%%%%%%%%%%
\section{14-band model}\label{sec:14}
%%%%%%%%%%%%%%%%%%%%%%%
\subsection{Energy spectrum and spin splittings in bulk semiconductor}
%%%%%%%%%%%%%%%%%%%%%%%

The 14-band $\bm k\cdot\bm p$ model, called sometimes 5-level $\bm k\cdot\bm p$ model or the 14$\times$14 extended Kane model, displays the full symmetry of a zinc-blende-lattice crystal and describes in detail the electron dispersion in the vicinity of the $\Gamma$ point in materials.\cite{winkler:book,Pikus1988,roessler,fasol,zaw,voon} The model includes the $\Gamma_{8v}$ and $\Gamma_{7v}$ valence bands formed from the orbital
Bloch functions $\mathcal X$, $\mathcal Y$, $\mathcal Z$ ($\Gamma_{15}$ representation in the coordinate system $x,y,z$), the lowest conduction band $\Gamma_{6c}$ formed from the invariant orbital function $\mathcal S$ ($\Gamma_1$ symmetry) and the higher conduction bands $\Gamma_{8c}$ and $\Gamma_{7c}$  originating from the $\Gamma_{15}$-symmetry orbital functions $\mathcal X'$, $\mathcal Y'$, $\mathcal Z'$.  For the spinor $\Gamma$-point Bloch functions $|N\rangle$ $(N=1\dots14)$, we use the notations 
\begin{equation} \label{basis1}
|\Gamma_{6c}, m\rangle, |\Gamma_{8v}, m'\rangle, |\Gamma_{7v}, m\rangle, |\Gamma_{8c}, m'\rangle, |\Gamma_{7c}, m\rangle\:, 
\end{equation}
where $m=\pm 1/2$ and $m'=\pm 1/2, \pm 3/2$, the $\Gamma_6$ basis is taken in the form $\uparrow$$\mathcal S$, $\downarrow$$\mathcal S$, and the $\Gamma_{8}$ basis is given by Eqs.~(\ref{psihh}) and (\ref{psilh}), the $\Gamma_7$ basic functions are also taken in the canonical form.\cite{Ivchenko} The model contains eight parameters of the 14-band model, namely, the band gap $E_g$, the energy distance $E_g'$ between the $\Gamma_{7c}$ and $\Gamma_{6}$ states, spin-orbit splittings $\Delta$ and $\Delta'$ of the valence and higher conduction bands, interband matrix elements of the momentum operator
\begin{eqnarray} \label{PPQ}
P &=& {\rm i}\frac{\hbar}{m_0} \braket{\mathcal S}{\hat{p}_x}{\mathcal X}\:,  \\
P^{\prime} &=& {\rm i}\frac{\hbar}{m_0} \braket{\mathcal S}{\hat{p}_x}{\mathcal X^{\prime}}\: ,\nonumber \\
Q &=& {\rm i}\frac{\hbar}{m_0} \braket{\mathcal X'}{\hat{p}_y}{\mathcal Z}\:, \nonumber
\end{eqnarray} 
and, finally, the interband matrix element of the spin-orbit interaction between the valence and higher conduction bands defined by
\begin{equation}
\label{Dm}
\Delta^- = 3 \langle \Gamma_{8c},m'|\mathcal H_{so} |\Gamma_{8v},m'\rangle = -\frac{3}{2}  \langle \Gamma_{7c},{m}|\mathcal H_{so} |\Gamma_{7v},{m}\rangle\:,
\end{equation}
where $\mathcal H_{so} =\hbar^2 ([\bm \sigma \times \nabla U (\bm r)]\bm p)/4m_0^2c^2$ is the spin-orbit Hamiltonian, $U({\bm r})$ is the  
spin-independent single-electron periodic potential, $c$ is the speed of light, and $m_0$ is the free-electron mass. Note that hereafter we ignore the difference between the generalized momentum operator ${\bm \pi}$ and the operator ${\bm p} = - {\rm i} \hbar {\bm \nabla}$ because the $\bm k\cdot({\bm \pi} - {\bm p})$ correction is usually negligibly small.\cite{liu} The 14$\times$14 Hamiltonian matrix ${\cal H}^{(14)}_{N'N}$ is a sum of the diagonal matrix elements $E^0_N \delta_{N'N}$ and off-diagonal matrix elements linear either in $\Delta^-$ or in ${\bm k}$. As frequently used in the simplified multiband $\bm k\cdot\bm p$ models,~\cite{winkler:book} we ignore the free-electron term $(\hbar^2 k^2/2 m_0)\delta_{N'N}$, which in the case of QW structures reduces the number of boundary conditions at an interface from 28 to 14 and simplifies numerical calculations.

\begin{table}[tb]
\caption{\label{tab:mass} Analytic expressions for the effective mass of an electron in the $\Gamma_6$ conduction band and the Luttinger parameters $\gamma_1$, $\gamma_2$ and $\gamma_3$ for the $\Gamma_{8v}$ band.\cite{winkler:book} }  
\begin{ruledtabular}
\begin{tabular}{ccc}
$\displaystyle \frac{m_0}{m_e}$          
 & $=$ &
  $ \displaystyle \frac{2m_0P^2}{\hbar^2 E_g}\left( 1- \frac13 \frac{\Delta}{E_g + \Delta} \right) - \frac{2m_0P'^2}{\hbar^2 E'_g}\left( 1- \frac23 \frac{\Delta'}{E'_g + \Delta'} \right)$\\
$\displaystyle \gamma_1$ & $=$ & $ \displaystyle\frac{2m_0}{3\hbar^2} \left(\frac{P^2}{ E_g} + \frac{Q^2}{E_g+E_g'} + \frac{Q^2}{E_g+E_g'+\Delta'}  \right)$  \\
$\displaystyle \gamma_2$ & $=$ &$ \displaystyle\frac{m_0}{3\hbar^2} \left(\frac{P^2}{ E_g} - \frac{Q^2}{E_g+E_g'} \right)$   \\
$\displaystyle \gamma_3$ &$=$ & $ \displaystyle\frac{m_0}{3\hbar^2} \left(\frac{P^2}{ E_g}  + \frac{Q^2}{E_g+E_g'} \right)$     
\end{tabular}
\end{ruledtabular}
\end{table}

The diagonalization of the 14-band Hamiltonian yields the electron energy spectrum in the bulk material. Table~\ref{tab:mass} summarizes four fundamental band parameters, the electron effective mass in the $\Gamma_6$ conduction band and the three Luttinger parameters for the $\Gamma_{8v}$ band, calculated in the second order of the $\bm k \cdot \bm p$ perturbation theory. Table~\ref{tab:parameters} shows three different parametrizations of the 14-band model used in Refs.~\onlinecite{Pikus1988}, \onlinecite{zaw}
and \onlinecite{jancu}. One can see that these parametrizations provide close values of the conduction-band effective mass and Luttinger parameters but as demonstrated below quite different values of ${\bm k}$-dependent spin-orbit splittings.

\begin{table*}[tb]
\caption{\label{tab:parameters} Three parametrizations of $\bm k \cdot \bm p$ model for GaAs used in the literature. Energy gaps are given in eV, matrix elements $P$, $P'$ and $Q$ are given in eV\AA. Conduction band effective mass and Luttinger parameters are calculated after expressions given in Tab.~\ref{tab:mass}.}  
\begin{ruledtabular}
\begin{tabular}{l|cccccccc|cccc}
 Parametrization                   & $E_g$ & $\Delta$& $E'_g$ & $\Delta'$ & $\Delta^-$ & $P$ & $P'$ & $Q$ & $m_e/m_0$ & $\gamma_1$ & $\gamma_2$ & $\gamma_3$ \\
                    \hline
Ref.~[\onlinecite{jancu}] (I) & 1.52 & 0.341 & 3.02 & 0.2 & -0.17& 9.88 & {0.41} & {8.68} & {0.063} & {8.51} & {2.08} & {3.53}\\
Ref.~[\onlinecite{Pikus1988}] (II) & 1.52 & 0.34 & 2.93 & 0.17 & -0.1& 10.3 & 3.3 & 6 & {0.062} & {7.51} & {2.7} & {3.4}\\
Ref.~[\onlinecite{zaw}] (III) & 1.52 & 0.341 & 2.97 & 0.17 & -0.061& 10.31 & 3 & 7.7 & {0.061} & {8.42} & {2.48} & {3.63}\\
\end{tabular}
\end{ruledtabular}
\end{table*}
%%%%%%%%%%%%

\begin{table*}[tb]
\caption{\label{tab:constants} The constants of spin-orbit splitting calculated after Eqs.~(\ref{cond:3rd})--(\ref{eq:dgv4}) (in eV\AA$^3$), corrections to the parameters $a_j$ calculated after Eq.~\eqref{das}, and spin-orbit constants for $\bm k\parallel [110]$ calculated after Eqs.~\eqref{gammahl} (in eV\AA$^3$) for three different parameterizations introduced in Tab.~\ref{tab:parameters}.}
\begin{ruledtabular}
\begin{tabular}{c|ccccc|ccc|cc}
Parametrization & $\gamma_{c0}$ & $\delta \gamma_c$& $ \gamma_c $ & $\gamma_{v0}$ & $\delta \gamma_v$ & $\delta a_1$ & $\delta a_2$ & $\delta a_3$ & $\gamma_{lh}$ & $\gamma_{hh}$ \\
\hline
(I) & {-2.39} & -22.0& -24.4 & {6.65} & 76.7 & 0.883 & -0.431 & -0.258  & 96.4  & 13.2 \\
(II) & -13.9 & -10.6& -24.5& 39.5 & 34.7 & 0.396 & -0.193 & -0.116 & 79.3 &  5.84 \\
(III) & -16.0 & -8.13 & -24.1 & 45.7 & 26.9 & 0.645 & -0.315 & -0.189 &  80.5 &  8.79
\end{tabular}
\end{ruledtabular}
\end{table*}

Following Ref.~[\onlinecite{Pikus1988}] we present the coefficients $\gamma_c$ and $\gamma_v$ in Eqs.~\eqref{eq:cond}
and \eqref{eq:G8SS} as sums $\gamma_c = \gamma_{c0} + \delta \gamma_c$ and $\gamma_v = \gamma_{v0} + a_1\delta \gamma_v /2$, respectively, which allows one to separate the  $\bm k \cdot \bm p$ third-order contributions ($\gamma_{c0}, \gamma_{v0}$) from the fourth order contributions ($\delta \gamma_c, \delta \gamma_v$), which include one order in $\Delta^-$. The third order contributions were found in Ref.~[\onlinecite{Pikus1988}] with the result 
\begin{eqnarray}
\label{cond:3rd}
\gamma_{c0} &=& -\frac43 P P' Q \frac{\Delta (E'_g+\Delta') + \Delta' E_g}{E_g E'_g(E_g + \Delta)(E'_g + \Delta')}\:,\\
\label{valence:3rd}
\gamma_{v0} &=& \frac43 P P' Q \frac{E_g + E'_g + \Delta'/2}{E_g(E_g+E'_g)(E_g + E'_g+\Delta')}.
\end{eqnarray}
Note that, as compared with Eqs.~(18) and~(20) of Ref.~[\onlinecite{Pikus1988}], we use here the different sign for $Q$, include the factor $\mathrm i \hbar/m_0$ in our definition of $P$ and $P'$, and refer $E_g'$ not to the $\Gamma_{8c}$ but to $\Gamma_{7c}$ band. 
The fourth-order correction to $\gamma_c$ reads
\begin{equation}
\label{cond:4rd}
\delta \gamma_c = \frac49 \Delta^{-}Q \frac{P^2(3E'_g + 2\Delta') + P'^2(3E_g + \Delta)}{E_g E'_g(E_g + \Delta) (E'_g + \Delta')}.
\end{equation}
The fourth-order corrections $\delta \gamma_v a_i /2 $ in the $\Gamma_8$ valence band are conveniently written as $(a_i^{(0)} + \delta a_i)\delta \gamma_v/2$ ($i= 1,2,3$), where
\begin{equation}
\label{eq:dgv4}
\delta\gamma_v = -\frac49 \frac{\Delta^- P^2 Q [{\Delta} + 2\Delta' + 3(E_g + E'_g)]}{\Delta E_g (E_g +E'_g + {\Delta})(\Delta' + E_g + E'_g)}\:.
\end{equation}
and 
\begin{equation}
\label{as}
a_{1}^{(0)}= \frac{13}{4}, \quad  a_{2}^{(0)}= -1, \quad a_{3}^{(0)}=1\:.
\end{equation}
The terms proportional to $a_i^{(0)}$ represent $\Delta^-P^2Q$ contribution, they were derived in Ref.~[\onlinecite{Pikus1988}]. Equation (\ref{eq:dgv4}) differs  from Eq.~(21a) in Ref.~[\onlinecite{Pikus1988}] by extra $\Delta$ in the numerator and denominator. In addition to $a^{(0)}_i$ there are other contributions proportional to $\Delta^- Q^3$ which are disregarded in Ref.~[\onlinecite{Pikus1988}] and can be presented in the form  
\begin{eqnarray}
\label{das}
\delta a_i =  c_i \frac{Q^2 E_g }{P^2(E_g + E'_g)}\:, \hspace{0.5 cm} \mbox{}\\
c_1 = \frac{41}{12}\:,\: c_2 =   - \frac53\:,\:c_3 =  - 1 \:. \nonumber 
\end{eqnarray}
These contributions may play an important role because $P$ and $Q$ are of the same order of magnitude. It is worth to mention that additional contributions $\propto \Delta^- P^{\prime 2}Q$ are negligibly small as compared to those in Eq.~(\ref{das}). 

For completeness, below we write down the expressions for the coefficients of the $k^3$ splittings $\Delta_{hh}(k) = \gamma_{hh} k^3$, $\Delta_{lh}(k) = \gamma_{lh} k^3$ of the heavy- and light-hole subbands for the particular direction $\bm k \parallel [110]$:
\begin{subequations}
\label{gammahl}
\begin{multline}
\label{gammah}
\gamma_{hh} = \frac12 \left| \gamma_v \left( 1 - \frac{3\xi - 1}{\sqrt{1 + 3\xi^2}} \right) + \delta \gamma_v \left[\left( 1 - \frac{2\xi}{\sqrt{1 + 3\xi^2}} \right)\right.\right.\\
+\frac12 \delta a_1 \left( 1 - \frac{3\xi - 1}{\sqrt{1 + 3\xi^2}} \right) +
 \frac18 \delta a_2 \left( 7 - \frac{21\xi - 13}{\sqrt{1+ 3\xi^2}} \right) \\ \left.\left. + \frac14 \delta a_3 \left( 1 + \frac{\xi}{\sqrt{1 + 3\xi^2}} \right) \right]\right|,
\end{multline}
\begin{multline}
\label{gammal}
\gamma_{lh} = \frac12 \left| \gamma_v \left( 1 + \frac{3\xi - 1}{\sqrt{1 + 3\xi^2}} \right) + \delta \gamma_v \left[ \left( 1 + \frac{2\xi}{\sqrt{1 + 3\xi^2}} \right)\right. \right.\\
 + 
\frac12 \delta a_1 \left( 1 + \frac{3\xi - 1}{\sqrt{1 + 3\xi^2}} \right) + 
  \frac18 \delta a_2 \left( 7 + \frac{21\xi - 13}{\sqrt{1+ 3\xi^2}} \right)\\ \left.\left. + \frac14 \delta a_3 \left( 1 - \frac{\xi}{\sqrt{1 + 3\xi^2}} \right) \right]\right|,
\end{multline}
\end{subequations}
where the ratio $\xi=\gamma_3/\gamma_2$ characterizes the valence band warping.
Table~\ref{tab:constants} summarizes the values of spin-orbit splitting constants for the conduction and valence bands calculated, again for different parameterizations of the $\bm k \cdot \bm p$ model. It is seen that the fourth-order contributions are comparable with and can be even larger than the third order ones. Moreover, the inclusion of the corrections $\delta a_i$ in Eqs.~\eqref{gammahl} significantly increases the spin splitting of heavy-hole states. For example, in the parametrization (I) (Ref.~\onlinecite{jancu}) the omission of $\delta a_i$ terms yields $\gamma_{hh} \approx 4.6$~eV\AA$^3$ while the corrected value is larger by almost a factor of 3.

\subsection{Boundary conditions and electronic states in QWs}

Let us now apply the 14-band model to calculate the energy spectrum in a symmetric QW grown along the $z \parallel [001]$ direction. In addition to the basis (\ref{basis1}), we use another set of basic functions $|l,s\rangle$ ($l=1\dots7, s= \pm 1/2$), where $|l,1/2\rangle = \uparrow$${\cal R}_l$ for the spin $s=1/2$ and $|l,-1/2\rangle = \downarrow$${\cal R}_l$ for  $s=-1/2$, and ${\cal R}_l$ are the orbital Bloch functions $\cal S, \cal X, \cal Y, \cal Z, \cal X', \cal Y', \cal Z'$. This basis consisting of products of the up- and down-spinors and the orbital functions is more convenient for the formulation and analysis of boundary conditions at the interfaces. Within the 14-band approach each electron state $\Psi_{n,j}$ in a QW is described by 14 envelope functions $f_{nj,ls}$ in the expansion
\begin{equation}
\label{eq:wfqw}
\Psi_{nj} = \frac{\mathrm{e}^{\mathrm i {\bm k}_{\parallel} {\bm \rho}}}{\sqrt{S}} \sum \limits_{ls} f_{nj,ls} (z) \ket{l,s}\:.
\end{equation}
Here $z$ is the growth axis, $\bm k_\parallel$ is the in-plane wave vector with two components $k_x,k_y$, $S$ is the normalization area, the subscript $n$ denotes the subband, e.g., $n = e1, hh1, lh1$ etc., and $j$ is a pseudospin index enumerating two states in each subband $n$ degenerated in the $\Gamma$-point ($\bm k_\parallel = 0$). The energy spectrum $E_{nj}(\bm k_\parallel)$ in the $n$-th electronic subband in $\bm k$-space is obtained from the numeric solution of the Schr\"{o}dinger equation
\begin{equation}
\label{eq:H14}
{\mathcal H}^{(14)}\left( \bm k_\parallel, \hat{k}_z \right) \Psi_{nj} = E_{nj} (\bm k_\parallel) \Psi_{nj}\:,
\end{equation}
where $k_z$ is replaced by a differential operator $\hat{k}_z = -\mathrm i \partial/\partial z$ acting on the envelopes $f_{nj,ls}(z)$. 

\begin{figure*}[htpb]
\includegraphics[width=0.9\textwidth]{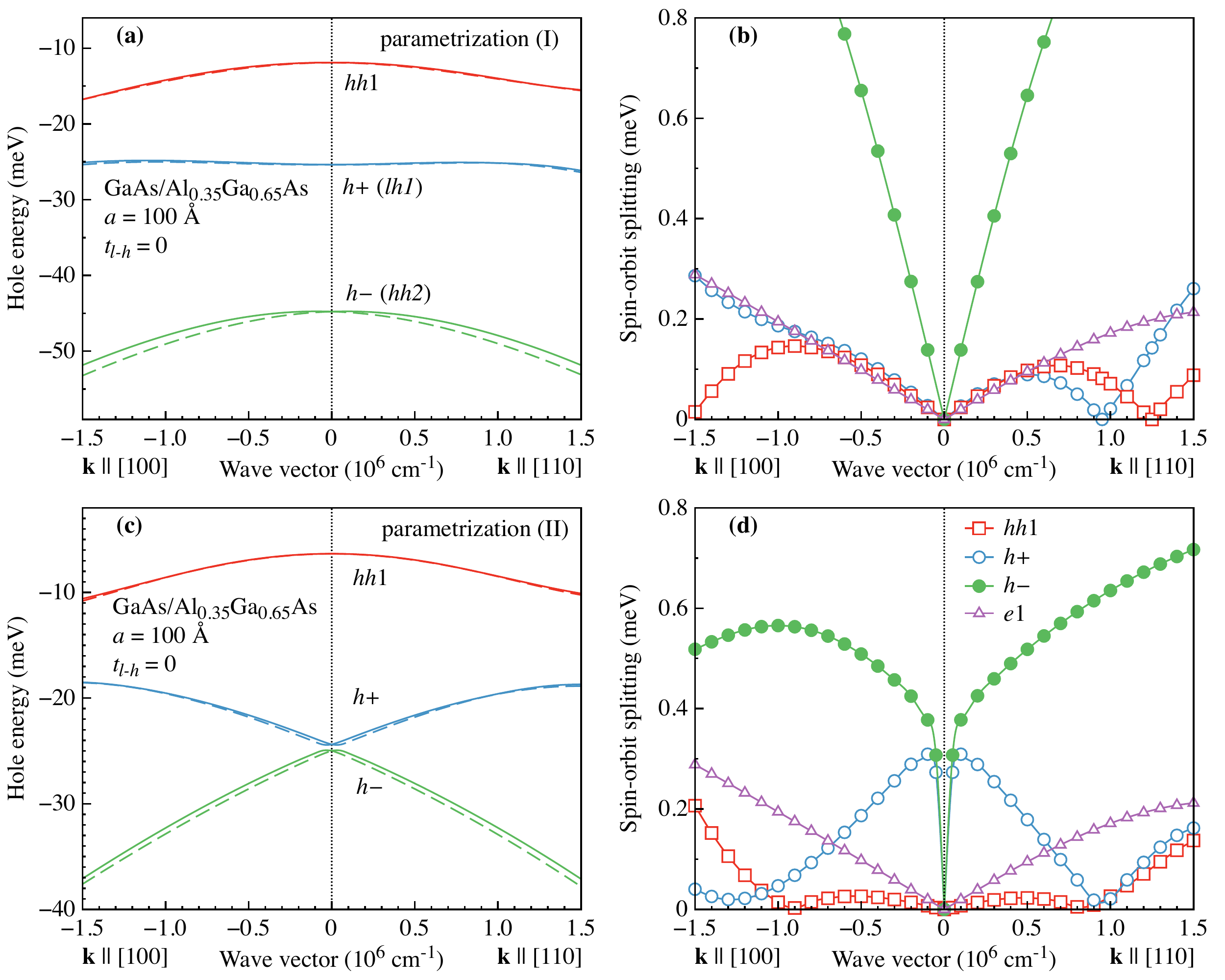}
\caption{\label{fig:fig1}
Dispersion (a, c) and spin splitting (b, d) of valence subbands for GaAs/Al$_{0.35}$Ga$_{0.65}$As QW. The calculations are done for two parametrizations (see Tab.~\ref{tab:parameters}): (I) [panels (a) and (b)] and (II) [panels (c) and (d)]. The spin splitting of conduction subband $e1$ is presented in (b) and (d) for comparison. 
}
\end{figure*}

Equation~\eqref{eq:H14} should be supplemented with the boundary conditions. The key requirement is associated with the conservation of the particle flux. For a 14-component envelope $f_{ls}$ the flux is given by
\begin{equation} \label{flux2}
{\bm S} = \frac{1}{\hbar} \sum\limits_{l's'ls} f^*_{l's'} \frac{\partial {\cal H}^{(14)}_{l's',ls}({\bm k})}{\partial {\bm k}} f_{ls}\:,
\end{equation}
or explicitly one has, e.g., for the flux $x$-component  
\begin{eqnarray} \label{flux}
S_x &=& \frac{\rm i}{\hbar} \left[ P \left( \hat{f}^{\dag}_{\mathcal X} \hat{f}_{\mathcal S} - \hat{f}^{\dag}_{\mathcal S} \hat{f}_{\mathcal X} \right) + P' \left( \hat{f}^{\dag}_{\mathcal X'} \hat{f}_{\mathcal S} - \hat{f}^{\dag}_{\mathcal S} \hat{f}_{\mathcal X'} \right) \right. \nonumber \\
&& \left. + Q \left(  \hat{f}^{\dag}_{\mathcal Z'} \hat{f}_{\mathcal X} - \hat{f}^{\dag}_{\mathcal X} \hat{f}_{\mathcal Z'} +  \hat{f}^{\dag}_{\mathcal X'} \hat{f}_{\mathcal Z} - \hat{f}^{\dag}_{\mathcal Z} \hat{f}_{\mathcal X'} \right) \right] \:,
\end{eqnarray}
where two-component spinor envelopes 
$$
\hat{f}_l = \left[ \begin{array}{c} f_{l,1/2}\\ f_{l,-1/2} \end{array} \right]
$$
are introduced for brevity.

Since the $\bm k \cdot \bm p$ Hamiltonian contains $k_z$ terms only of the first order, it is enough to impose one condition per envelope $f_{nj,ls}$. In what follows we assume that the interband matrix elements $P$, $P'$, $Q$, and $\Delta^-$ are the same in the QW and barrier materials, so that solely the diagonal elements $E^0_N$, i.e. the positions of bands at $\bm k$ = 0, experience discontinuities at heterointerfaces. In this case the simplest and intuitively natural set of boundary conditions conserving the flux could be merely the continuity of all envelopes at the interface, see Eq.~(\ref{flux}). 
However, such boundary conditions do not account for the reduced microscopic symmetry of a single interface described by the $\mathrm C_{2v}$ point group and caused by the anisotropy of chemical bonds in the (001) plane.\cite{aleiner92eng,ivchenko96,krebs96,NestokGolub} Therefore, we are interested in boundary conditions as simple as possible but those which conserve the flux and make allowance for the interface heavy-light hole mixing. We recall that in calculations based on 4$\times$4 Luttinger Hamiltonian and four-component envelope $\Phi$ this kind of state mixing is described by an extra term in the boundary conditions\cite{ivchenko96}
\begin{eqnarray} \label{lhmixing}
\Phi_A = \Phi_B\:, \hspace{3 cm} \mbox{} \\
\hat{M}^{-1}_A \left( \frac{\partial \Phi}{\partial z} \right)_A =
\hat{M}^{-1}_B \left( \frac{\partial \Phi}{\partial z} \right)_B + \frac{2}{\sqrt{3}} \frac{t_{l\mbox{-}h} }{a_0 m_0} \{ J_x J_y\}_s \Phi\:, \nonumber
\end{eqnarray}
where the matrix $M$ is diagonal and comprises the values of heavy-hole, $m_{hh} = m_0/(\gamma_1 - 2 \gamma_2)$, and light-hole,
$m_{lh} = m_0/(\gamma_1 + 2 \gamma_2)$, effective masses in the [001] direction, $t_{l\mbox{-}h}$ is a real coefficient, and $a_0$ is the lattice constant. 
In order to include the hole mixing effect into the 14-band $\bm k \cdot \bm p$ model, we provide the minimal generalization of natural boundary conditions for envelopes $f_{nj,ls}$ as the requirement of continuity of the five envelopes $f_{nj,ls}$ corresponding to ${\cal R}_l = \mathcal S, \mathcal X, \mathcal Y, \mathcal Z$ and $\mathcal Z'$ and the following discontinuity of the remaining envelopes, as follows, 
\begin{eqnarray}
\label{eq:BC_int}
\left(\hat{f}_{nj,\cal X'}\right)_A  &=& \left(\hat{f}_{nj,\cal X'} \right)_B + \tilde{t} \left( \hat{f}_{nj,\cal X}  \right)_B\:, \\
\left(\hat{f}_{nj,\cal Y'} \right)_A  &=& \left(\hat{f}_{nj,\cal Y'} \right)_B + \tilde{t} \left(\hat{f}_{nj,\cal Y} \right)_B\:, \nonumber 
\end{eqnarray}
where $\tilde{t}$ is a real dimensionless interface-mixing parameter. One can readily see that the proposed conditions (\ref{eq:BC_int}) are in agreement with the flux continuity. The boundary conditions for the $4\times 4$ model can be obtained from Eqs.~\eqref{eq:BC_int} taking into account that the $Q k_z$ off-diagonal matrix elements in the 14$\times$14 Hamiltonian couple $\hat{f}_{nj,\cal X'}$ with $\hat{f}_{nj,\cal Y}$ and $\hat{f}_{nj,\cal Y'}$ with $\hat{f}_{nj,\cal X}$. As a result we arrive in the linear-$k_z$ approximation at the second boundary condition (\ref{lhmixing})
with the heavy-light hole mixing coefficient
\begin{equation}
\label{relation}
t_{l\mbox{-}h} \equiv \frac{2 m_0 a_0}{\sqrt{3}\hbar^2}Q\tilde{t} \:.
\end{equation}
In what follows we use $t_{l\mbox{-}h}$ as an independent parameter of our theory. It is worthwhile to stress that an inclusion of other extra terms in the 14-band boundary conditions modifies Eqs. (\ref{lhmixing}) to a more complicated form of Ref.~\onlinecite{glinskii}.

\begin{figure*}[htpb]
\includegraphics[width=0.9\textwidth]{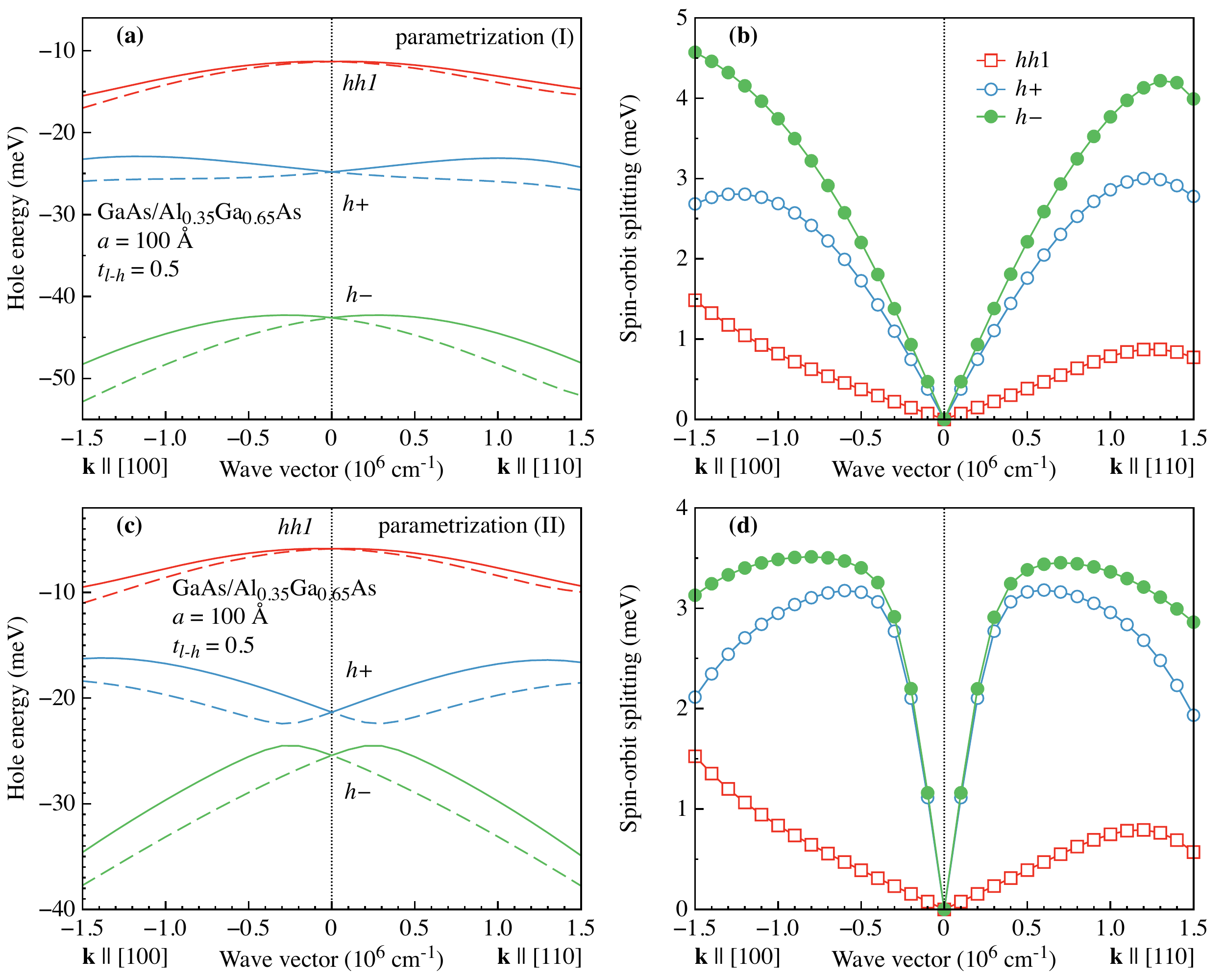}
\caption{\label{fig:fig2}
Dispersion (a, c) and spin splitting (b, d) of valence subbands for GaAs/Al$_{0.35}$Ga$_{0.65}$As QW with account for the interface mixing of heavy and light holes [$t_{l\mbox{-}h} = 0.5$ corresponding to $\tilde{t} \approx 0.07$ for the set (I) and $\tilde{t} \approx 0.1$ for the set (II)]. The calculations are done for two parametrizations (see Tab.~\ref{tab:parameters}): (I) [panels (a) and (b)] and (II) [panels (c) and (d)].
}
\end{figure*}

%%%%%%%%%%%%%%%%%%%%%%%%%%%%%%%%%%%%%%%%%%%
\section{Results and discussion}\label{sec:res}
%%%%%%%%%%%%%%%%%%%%%%%%%%%%%%%%%%%%%%%%%%%
Figures \ref{fig:fig1} and \ref{fig:fig2} display the main results of our 14-band model calculation, with the generalized boundary conditions \eqref{eq:BC_int}, of valence band structure in a 100-\AA-thick GaAs/Al$_{0.35}$Ga$_{0.65}$As QW. The energy dispersion of three topmost valence subbands is obtained neglecting, Fig.~\ref{fig:fig1}, and taking into account, Fig.~\ref{fig:fig2}, the heavy-light hole interface mixing. Each figure contains four panels. Panels (a), (b) are calculated for the parametrization (I), see Tab.~\ref{tab:parameters}, panels (c), (d) represent the parametrization~(II). Solid and dashed lines in panels (a), (c) show the subband dispersion, while the wave vector dependence of the spin splittings is depicted in panels (b) and (d). As mentioned above we assume the same values of $P$, $P'$, $Q$, $\Delta$, $\Delta'$ and $\Delta^-$ for the well and barrier materials, the band gap of Al$_x$Ga$_{1-x}$As solid solution is taken from the quadratic equation\cite{kiselev} $E_g(x) = E_g(0) + 1.04x + 0.46x^2$. The standard ratio 2/3 is used for the  $\Gamma_{8v}$- and $\Gamma_{6c}$-band off-sets, $\Delta E_v$ and $\Delta E_c$. The off-set $\Delta E_{c'}$ for the higher conduction band is chosen to equal $-\Delta E_v$ since all proposed parametrizations give practically the same value for the sum $E_g + E_g'$ in GaAs and AlAs. 

We begin the discussion from comparison of the $\Gamma$-point positions of the valence subbands calculated by using the 14-band model and 4$\times$4 Luttinger Hamiltonian in the absence of interface mixing, $t_{l\mbox{-}h}=0$. In the four-band Luttinger model the heavy- and light-hole envelope functions and $\Gamma$-point energies $E_{hh\nu}$, $E_{lh\nu}$ ($\nu=1,2\dots$) are found from 
\begin{eqnarray} \label{Lutt}
\hspace{8 mm}\left[ -\frac{\hbar^2}{2} \dz \frac{1}{m_{hh}} \dz + V(z)\right] \Phi_{hh\nu} = E_{hh\nu} \Phi_{hh\nu}\:, \nonumber\\
\left[ -\frac{\hbar^2}{2} \dz \frac{1}{m_{lh}} \dz + V(z)\right] \Phi_{lh\nu} = E_{lh\nu} \Phi_{lh\nu}\:, \hspace{3 mm} \mbox{}
\end{eqnarray}
where $V(z)$ is the confinement potential determined by the valence band offsets, and functions $\Phi(z)$ satisfy the Bastard boundary conditions given by the Eqs.~(\ref{eq:BC_int}) at $t_{l\mbox{-}h}=0$. It turns out that the $\Gamma$-point energies calculated in the 14-band and Luttinger models agree with each other within 5\% accuracy. 
It is seen from Figs.~\ref{fig:fig1}(a) and \ref{fig:fig1}(c) that the two sets of parameters result in significantly different positions of the two lower subbands at $\bm k_\parallel =0$ despite relatively small ($\lesssim 30\%$) difference of the Luttinger parameters. For the parameter set (II), these subbands are much closer in energy than for the set (I). Moreover, the calculation carried out within the Luttinger Hamiltonian model for the parametrization (II) gives the crossing between the $hh2$ and $lh1$ states at $a\approx95$~\AA,~see Ref.~\onlinecite{durnev} for details. For the 100~\AA-thick QW, the pure-state energies still lie very close to each other, the mixing is remarkable and we use for these $\Gamma$ states in Figs.~\ref{fig:fig1} and \ref{fig:fig2} the notation $h+,h-$ instead of $lh1$ and $hh2$, respectively.

The heavy-light hole mixing is crucial both for the $h_\pm$ states and the spin-orbit splitting of the valence subbands. In the 4$\times$4 effective Hamiltonian approach one could expect to get the $hh2$-$lh1$ mixing by including the $V_z \hat{k}^3_z$ term of Eq.~(\ref{eq:G8SS}). The matrix $V_z$ indeed mixes the $|\Gamma_8, 3/2\rangle$ state with the $|\Gamma_8,-1/2\rangle$
state as well as $|\Gamma_8, -3/2\rangle$  with $|\Gamma_8,1/2\rangle$. The inclusion of the last term in Eq.~(\ref{eq:G8SS}) into the effective Hamiltonian makes however the problem unsolvable, neither rigorously nor approximately, even for the infinitely high barriers in which case the boundary conditions reduce to vanishing of the envelopes at the both interfaces. Firstly, strictly speaking, the order of differential equations increases, and additional unphysical solutions appear.  Secondly, an attempt to take the operator $V_z \hat{k}^3_z$ into account as a perturbation encounters the problem of a nonhermitian nature of the $\hat{k}^3_z$ operator, $\langle hh2 | \hat{k}^3_z | lh1 \rangle \neq \langle lh1 | \hat{k}^3_z | hh2 \rangle^*$, defined on the space of envelopes satisfying Eqs.~(\ref{Lutt}) and vanishing at the interfaces. It is worth to mention that, due to the special reason,\cite{note:reason} the linear-${\bm k}$ terms in the valence subbands $hh\nu,lh\nu$ related to the $V_z \hat{k}^3_z$ operator and found as a first-order correction to the energy spectrum in Ref.~[\onlinecite{Rashba1988175}] are finite and can be compared with the 14-band calculations, see below. For barriers of finite height the inconsistency related to the $V_z \hat{k}^3_z$ operator seems insurmountable for finding both the $\Gamma$-point energies and the linear-${\bm k}$ dispersion. On the other hand, the 14-band $\bm k\cdot \bm p$ model under study allows one to derive the cubic terms of Eq.~(\ref{eq:G8SS}) for bulk materials and to compute the QW valence eigenstates comprising an admixture of the Bloch functions  $|\Gamma_8, 3/2\rangle$ and $|\Gamma_8,-1/2\rangle$ or $|\Gamma_8, -3/2\rangle$  with $|\Gamma_8,1/2\rangle$ at the point $k_x=k_y=0$.

The curves in Fig.~\ref{fig:fig2} are calculated for the same set of parameters as in the previous figure, with one exception: Now the interface mixing parameter $\tilde{t}$ in the boundary conditions (\ref{eq:BC_int}) is nonzero and corresponds to a reasonable value of the parameter $t_{l\mbox{-}h}=0.5$ related with $\tilde{t}$ by Eq.~(\ref{relation}). Comparing Figs.~\ref{fig:fig1} and \ref{fig:fig2} we observe striking effects of the interface mixing. The splitting between the $h+$ and $h-$ states in Fig.~\ref{fig:fig2}(c) tremendously increases. Furthermore, all the spin splittings in Fig.~\ref{fig:fig2} are enhanced by about an order of magnitude. The positions of 
the $h+$ and $h-$ states at the $\Gamma$-point can perfectly be evaluated in the framework of Luttinger Hamiltonian and generalized boundary conditions \eqref{lhmixing}. In the first order in $t_{l\mbox{-}h}$ the role of extra term in the boundary conditions \eqref{lhmixing} can be reduced to an effective matrix element 
\begin{equation}
\label{Dlh:i}
\Delta_{l\mbox{-}h} = \frac{t_{l\mbox{-}h} \hbar^2}{m_0 a_0} \Phi_{hh2}(z_i)\Phi_{lh1}(z_i)
\end{equation}
that mixes the $lh1$ and $hh2$ states at $\bm k_\parallel =0$. Here $z_i$ is the coordinate of the right-hand interface. The mixed state energies are given by
\begin{equation}
\label{pm}
E_{h\pm} = \frac{E_{hh2} + E_{lh1}}{2} \pm \sqrt{\left( \frac{E_{hh2} - E_{lh1}}{2} \right)^2 + \Delta_{l\mbox{-}h}^2} \:.
\end{equation}
At the crossing point, where $E_{hh2} = E_{lh1}$, the splitting between the $h+$ and $h-$ eigenstates equals $2 |\Delta_{l\mbox{-}h}|$
and each of them is an equal admixture of the $hh2$ and $lh1$ pure states. The comparison of Figs.~\ref{fig:fig1}(c) and \ref{fig:fig2}(c) shows that, in parametrization II, the ``bulk'' $hh2$-$lh1$ mixing inside the QW layer is by an order of magnitude smaller than that due to the interface effect.
 
\begin{figure}[htpb]
\includegraphics[width=0.45\textwidth]{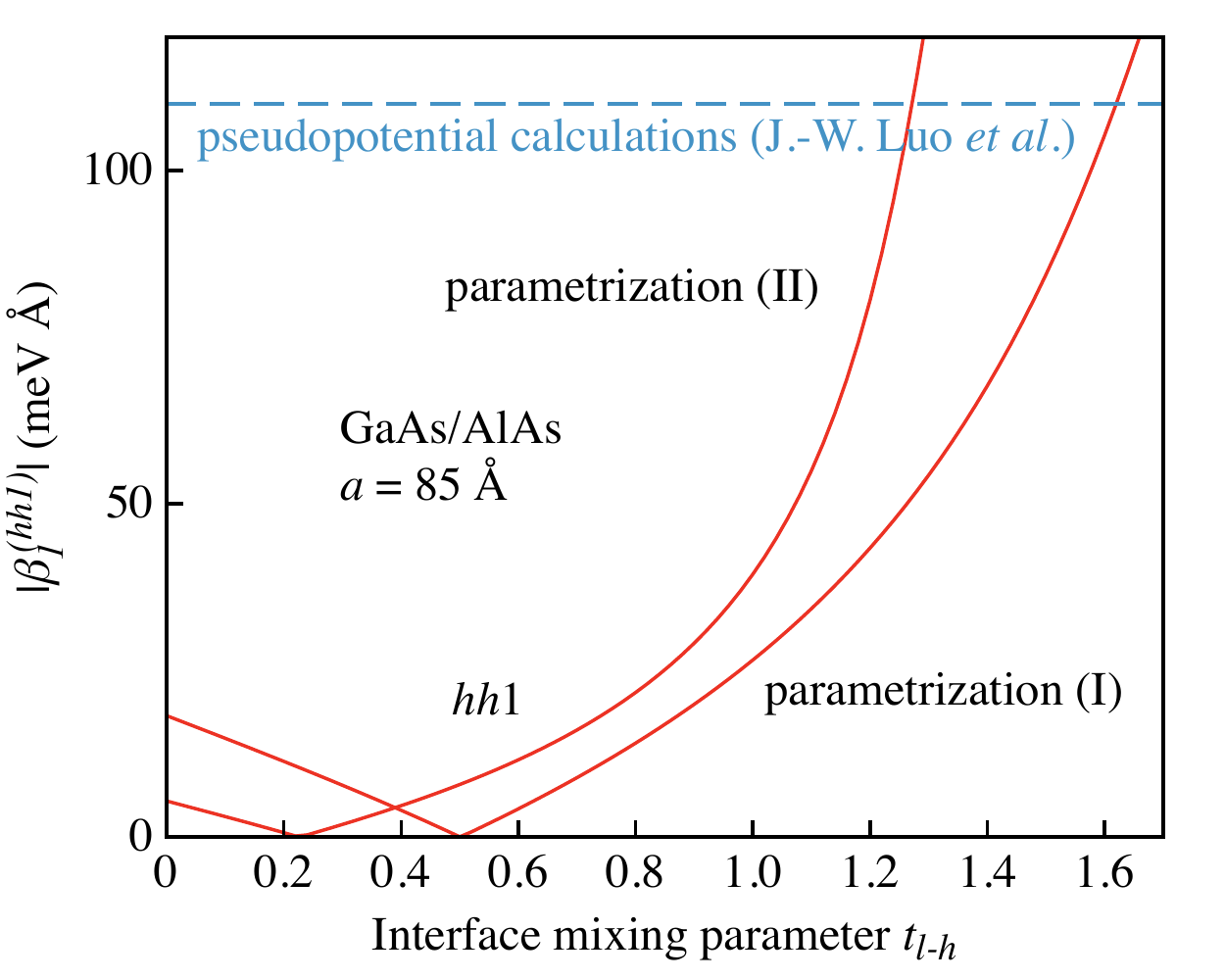}
\caption{\label{fig:fig3} The results of calculations for heavy-hole ($hh1$) spin splitting as a function of the interface mixing strength for a GaAs/AlAs 85~\AA~well. The dashed line indicates the result of pseudo potential calculations obtained for the same QW in Ref.~\onlinecite{PhysRevLett.104.066405}.
}
\end{figure}

\begin{figure*}[tb]
\includegraphics[width=0.95\textwidth]{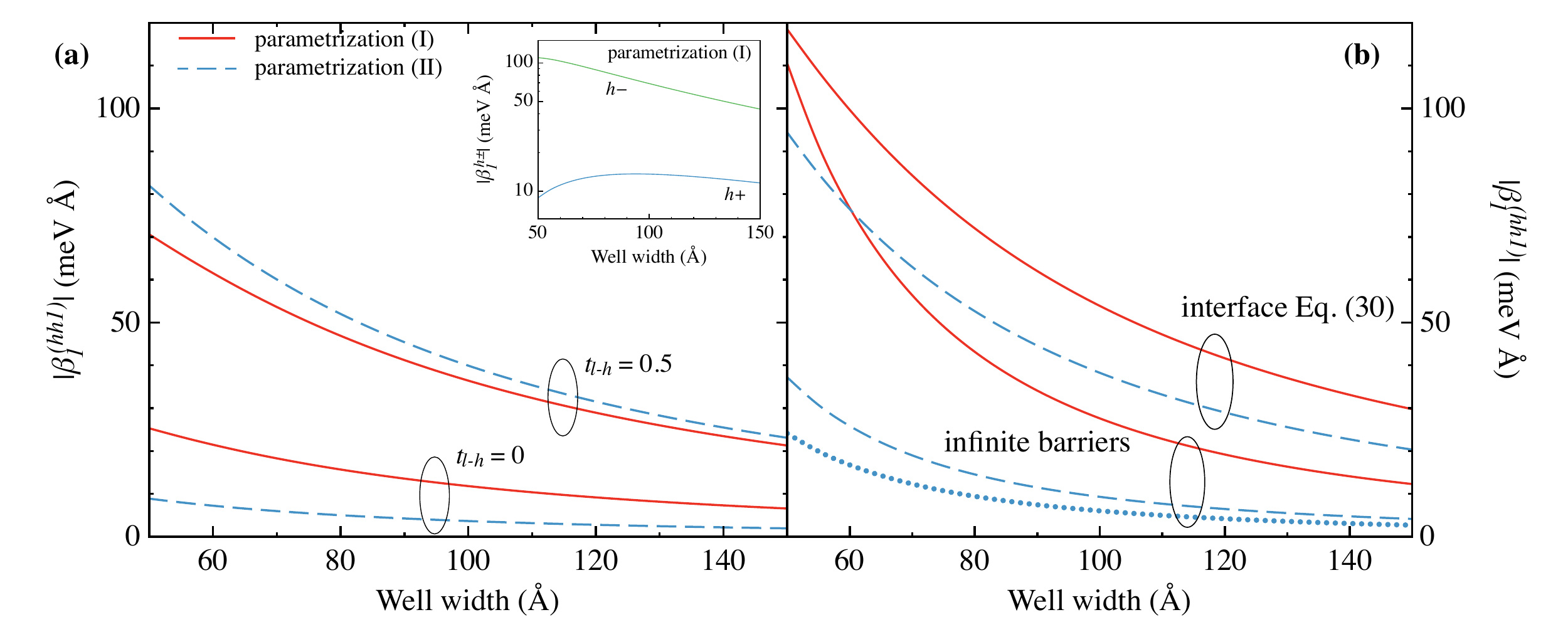}
\caption{\label{fig:fig4} Spin-orbit ${\bm k}$-linear term $\beta_1^{(n)}$ for the $hh1$ subband in a GaAs/Al$_{0.35}$Ga$_{0.65}$As QW. (a) 14-band numerical calculation is shown for two sets of parameters (solid and dashed lines) and for two values of interface mixing parameter: $t_{l\mbox{-}h} = 0$ and $t_{l\mbox{-}h} = 0.5$. The inset represents the results for $h+$ and $h-$ subbands at $t_{l\mbox{-}h} = 0$ for the parameterization (I).
(b) Analytical calculation of $\beta_1^{(hh1)}$. Three bottom curves are obtained in the limit of infinitely-high barriers from Eq.~(8) of Ref.~\onlinecite{Rashba1988175}: the solid curve represents the parametrization (I), the dotted and dashed curves are calculated for the parametrization (II) neglecting and taking into account the corrections $\delta a_i$ in Eq.~\eqref{das}, respectively. Two top curves demonstrate the interface-induced spin splitting according to Eq.~\eqref{eq:beta_hh1} with $t_{l\mbox{-}h} = 0.5$.}
\end{figure*}

The 14-band model developed here automatically provides the spin-orbit splitting of conduction and valence subbands for $\bm k_\parallel \ne 0$. Like the energy dispersion presented in panels (a,c), the spin splittings are given for two directions of the in-plane wavevector $\bm{k}_{\parallel} \parallel [100]$ and $\bm{k}_{\parallel} \parallel [110]$. It is noteworthy that the anisotropy of spin splittings becomes pronounced at $k_\parallel \sim 10^{6}$~cm$^{-2}$.  One can see in Figs.~\ref{fig:fig1}(b,d) that, even in the absence of interface mixing, the $\bm k$-linear spin splitting of $hh1$ subband is comparable with that for $e1$ conduction subband also shown (by triangles). The other notable feature is a huge linear-in-$\bm k$ spin splitting of $hh2$ and $lh1$ (or $h+$ and $h-$) subbands which is particularly pronounced for the set of parameters (II) where these states are close in energy. For this set of parameters the linear terms for $h_\pm$ states markedly exceed those for $e1$ electron and $hh1$ valence subbands. Such a behavior was uncovered by Rashba and Sherman\cite{Rashba1988175} in the model of infinite barriers: It is caused by (i) the heavy-light-hole mixing by the off-diagonal elements of Luttinger Hamiltonian proportional to $\hat k_z (k_x \pm \mathrm i k_y)$  and (ii) $V_z \hat{k}_z^3$ term in the bulk spin-orbit Hamiltonian~\eqref{eq:G8SS}.  The heavy-light-hole interface mixing results in the considerable enhancement of spin splittings of the valence subbands but has only a small effect on the spin splitting in the $e1$ conduction subband.\cite{keinz} 

To analyze further the effect of interface mixing, we present in Fig.~\ref{fig:fig3} the absolute value $|\beta_1^{(hh1)}|$ as a function of $t_{l\mbox{-}h}$ for a $85$~\AA-thick~GaAs/AlAs QW. The spin splitting constant $\beta_1^{(hh1)}$ in Eq.~\eqref{100} vanishes at a particular value of $t_{l\mbox{-}h} \approx$ 0.2 either 0.5 for the parametrizations (II) and (I), respectively, where the bulk-inversion asymmetry and interface-inversion asymmetry contributions to the splitting cancel each other. Here, it is worth to remark that the cancellation takes place for positive values of $t_{l\mbox{-}h}$, while for $t_{l-h}<0$ the absolute value of the splitting monotonously increases with an increase of $|t_{l-h}|$. The sign of this parameter lies beyond the limits of this work. Atomistic calculation of the spin-orbit splitting performed in Ref.~[\onlinecite{PhysRevLett.104.066405}] predicted huge values of $\beta_1^{(hh1)}$ for the $hh1$ subband reaching $115$~meV\AA~for a GaAs/AlAs structure with the GaAs layer thickness of $85$~\AA, as shown in Fig.~\ref{fig:fig3} by the horizontal dashed line. For this particular structure the 14-band model yields the same value of $hh1$ spin-orbit splitting assuming $t_{l\mbox{-}h} = 1.2 \div 1.6$. Relatively high values of interface-mixing parameter meet the expectation of a monotonous increase of $t_{l\mbox{-}h}$ with the content $x$ of the heteropair GaAs/Al$_x$Ga$_{1-x}$As. The dependence of $\beta_1^{(hh1)}$ on the QW width $a$ is shown in Fig.~\ref{fig:fig4}(a). One can see that for the both sets of parameters the interface mixing dominantly contributes to the spin splitting. In the selected well-width range, $\beta_1^{(hh1)}$ is a monotonous function of $a$. For smaller values of the width this coefficient reaches a maximum and then decreases with the increasing penetration of the wavefunction into the barriers. For completeness, the variation of spin-splitting coefficients $\beta_1^{(h\pm)}$ with $a$ in the $h\pm$ subbands is included in Fig.~\ref{fig:fig4}(a), see the inset. 

The numerical results presented above can be interpreted in terms of three independent contributions to the spin-orbit splitting of the valence subbands. First one is similar to that in the conduction band and, for the heavy-hole subband $hh1$, originates from the $\sum_\alpha J_\alpha^3 \varkappa_\alpha$ and $\left( V_xk_x + V_yk_y \right)k_z^2$ terms in the spin-orbit Hamiltonian for the $\Gamma_8$ band, Eq.~\eqref{eq:G8SS}, averaged over the size-quantization wavefunction. The second contribution results from the interference of the $\propto V_z k_z^3$ term in Eq.~\eqref{eq:G8SS} and off-diagonal elements $H$ of the Luttinger Hamiltonian. In evaluation of this second contribution one encounters the ``dangerous'' $k_z^3$ matrix element which can be calculated only in the limit of infinite barriers. In this limit, the sum of two contributions is given by Eq.~(8) of Ref.~\onlinecite{Rashba1988175}.
The third contribution to the $k$-linear splitting of the heavy-hole subband arises from the interface-induced heavy-light-hole mixing and becomes dominant for $|t_{l\mbox{-}h}| \gtrsim 1$. It can be evaluated within the 4-band model using the Luttinger Hamiltonian and the boundary conditions Eq.~\eqref{lhmixing} and taking into account that for $t_{l\mbox{-}h} =0$ the heavy-hole wave functions can be presented as~\cite{merkulov}
\[
\Psi_{\pm 3/2} = \Phi_{hh1}(z)| \Gamma_8,\pm 3/2\rangle \pm \mathrm i (k_x \pm \mathrm i k_y) S_{lh}(z)| \Gamma_8,\pm 1/2\rangle\:.
\]
Here the admixture of $|\Gamma_8,\pm 1/2\rangle$ states is considered in the first order in $\bm k_\parallel$, and the function $S_{lh}$ is found from~\cite{durnev}
\begin{multline}
\label{eq:S}
\left[ -\frac{\hbar^2}{2} \dz \frac{1}{m_{lh}} \dz + V(z) - E_{hh1} \right] S_{lh}(z) = \\
= - \frac{\sqrt{3} \hbar^2 }{m_0 a_0} \left\{ \gamma_3 \dz \right\}_s \Phi_{hh1}(z)\:.
\end{multline}
Here $E_{hh1}$ is the energy of $hh1$ subband in the $\Gamma$-point, Eq.~\eqref{Lutt}, and as before the curly brackets assume symmetrization of operators. Allowance for the $t_{l\mbox{-}h}\ne 0$ in Eq.~\eqref{lhmixing} gives rise to the interface inversion asymmetry contribution to the $hh1$ subband, which in the first order in heavy-light hole interface mixing reads~\cite{Golub, Vervoort}
\begin{equation}
\label{eq:beta_hh1}
 \beta_{1;{\rm int}}^{(hh1)}  = \frac{2 t_{l\mbox{-}h} \hbar^2}{m_0 a_0} a \Phi_{hh1}(z_i)S_{lh}(z_i)\:.
\end{equation}

The results of analytical calculations of the above contributions to the $hh1$ spin splitting are presented in Fig.~\ref{fig:fig4}(b). The two sets of curves are depicted: the set of three bottom curves corresponds to the ``bulk'' contribution calculated in the limit of infinite-barrier well, Ref.~\onlinecite{Rashba1988175}, and the set of two top curves represents the interface-induced contribution calculated after Eq.~\eqref{eq:beta_hh1}, for $t_{l\mbox{-}h} = 0.5$ and the finite barriers corresponding to a GaAs/Al$_{0.35}$Ga$_{0.65}$As QW. For calculation of the ``bulk'' contribution, bottom dashed and solid curves, we used parameters $a_i$ with inclusion of corrections Eq.~\eqref{das}. For comparison, the dotted curve in Fig.~\ref{fig:fig4}(b) is calculated for $a_i = a_i^{(0)}$ according to Eq.~(8) of Ref.~\onlinecite{Rashba1988175} for the set of parameters (II). From the bottom curves in Fig.~\ref{fig:fig4}(a) and Fig.~\ref{fig:fig4}(b) it is clearly seen that the infinite-barrier model strongly overestimates the ``bulk'' contribution to the spin splitting found within the 14-band model for $t_{l\mbox{-}h} = 0$. This can be attributed mainly to (i) significantly larger values of the $\hat{k}_z^2$ operator averaged over the heavy-hole envelope $\Phi_{hh1}$ found in the infinite-barrier well compared to the case of finite barriers, and (ii) the overestimation of $V_zk_z^3$ effect.

Similar analytical procedure can also be used to calculate the interface induced $\bm k$-linear spin-orbit splitting of the $h_\pm$ subbands. The detailed discussion of the energy spectrum for this pair of subbands will be presented elsewhere, here we resort to a simple resonant approximation which neglects all energy bands but $h_+$ and $h_-$. In this case, the dominant contribution results from the interface-induced mixing of the heavy- and light- hole states and reads~\cite{Toropov}
\begin{equation}
\label{hpm}
\beta_1^{(h_\pm)} = \pm \frac{{2}\sqrt{3} \hbar^2}{m_0}\frac{\Delta_{l\mbox{-}h} \left| \braket{lh1}{\left\{ \gamma_3 \hat{k}_z \right \}_s}{hh2} \right|}{\sqrt{(E_{hh2} - E_{lh1})^2 + {4\Delta_{l\mbox{-}h}^2}}}\:,
\end{equation}
\[
 \braket{lh1}{\left\{ \gamma_3 \hat{k}_z \right \}_s}{hh2} = \int \Phi_{lh1}(z) {\left\{ \gamma_3 \hat{k}_z \right \}_s} \Phi_{hh2}(z) dz\:.
\] 
Equation~\eqref{hpm} closely reproduces the results of numerical calculation of $\beta_1^{(h_\pm)}$ for parametrization (II)  where the subbands $h_\pm$ are particularly close in energy.

Above we have paid the main attention to the ${\bm k}$-linear spin splitting of valence subbands. However, it follows from the 14-band calculations presented in Fig.~\ref{fig:fig1}(b, d) and Fig.~\ref{fig:fig2}(b, d) that, at $k_\parallel \sim 10^{6}$~cm$^{-1}$, cubic in ${\bm k}$ terms begin to play essential role resulting in the anisotropy of the spin splitting. The $k^3$ contribution of $n$-th hole subband in $[001]$-grown QWs contains two independent parameters $\gamma_1^{(n)}$ and $\gamma_2^{(n)}$~\cite{cartoixa}
\begin{multline}
\label{eq:Ham_cubic}
\mathcal H_{n}^{[001]} = \gamma_1^{(n)} \left( \sigma_{x_1} k_{y_1}^3 + \sigma_{y_1} k_{x_1}^3 \right) \\ + \gamma_2^{(n)} \left( \sigma_{x_1} k_{x_1}^2 k_{y_1} + \sigma_{y_1} k_{y_1}^2 k_{x_1} \right)\:.
\end{multline}
The parameters of Hamiltonian Eq.~\eqref{eq:Ham_cubic} extracted from numerical simulation of a 100\AA-GaAs/Al$_{0.35}$Ga$_{0.65}$As QW are listed in Tab.~\ref{tab:cubic}. It is worth to stress that (i) interface mixing makes a significant contribution both to $k$-linear and $k^3$ terms in the valence band effective Hamiltonian, and (ii) the $k$-linear terms given by $\beta_1^{(n)}$ indeed exceed by far the $k$-linear terms in the bulk valence-band Hamiltonian.

\begin{table}[tb]
\caption{\label{tab:cubic} Valence-band spin splittings for a 100~\AA~GaAs/Al$_{0.35}$Ga$_{0.65}$As QW.}  
\begin{ruledtabular}
\begin{tabular}{llccc}
 &$n$ & $\beta_1^{(n)}$(meV\AA) & $\gamma_1^{(n)}$ (eV\AA$^3$) & $\gamma_2^{(n)}$ (eV\AA$^3$)\\
\hline
$t_{l\mbox{-}h} = 0$ & $hh1$ & 12.2 & $-$82 & $-$31 \\
 & $h+$ & 13.5 & $-$153 & 54 \\
 & $h-$ & 67 & $-$140 & $-$78 \\ 
\hline
$t_{l\mbox{-}h} = 0.5$ & $hh1$ & 36 & 55 & 29 \\
 & $h+$ & 186 & $-$412 & $-$677 \\
 & $h-$ & 230 & $-$475 & $-$475 \\ 
\end{tabular}
\end{ruledtabular}
\end{table}

\section{Conclusions}\label{sec:concl}

To conclude, we have presented here the 14-band $\bm k\cdot \bm p$ model extended to allow for the reduced microscopic symmetry of QW interfaces which makes it possible to calculate the spin-orbit splitting of hole subbands in QWs. We proposed a simple boundary condition, Eq.~\eqref{eq:BC_int}, which takes into account heavy-light hole mixing at the interface due to anisotropic orientation of interface chemical bonds. Main contributions to the hole spin splitting are identified. The developed model has been applied to calculate the valence-band spin splittings in (001) QWs, but it can be used as well for QWs of any crystallographic orientation including the (110) and (111) orientations. The results of numerical calculations are well described by the developed analytical theory. Moreover, we have demonstrated that the large values of the spin splitting for the topmost heavy-hole subband predicted in Ref.~\onlinecite{PhysRevLett.104.066405} on the basis of atomistic calculations can be ascribed to the relatively strong interface-induced mixing of heavy- and light-hole states.

\acknowledgments
We are grateful to M.O. Nestoklon and E.Ya. Sherman for valuable discussions.

This work was supported by RFBR, Dynasty Foundation, as well as EU project POLAPHEN.


\begin{thebibliography}{99}
\bibitem{note:reason} The spin-orbit splitting of the $hh1$ subband contains symmetrized combinations $\langle hh1 |k_z^2|lh\nu\rangle^*+\langle lh\nu |k_z^2|hh1\rangle$ which are meaningful for the envelope functions found in the limit of infinite barriers.
\end{thebibliography}
\end{document}